\documentclass[journal,twoside,web]{ieeecolor}
\usepackage{tuffc}
\usepackage{cite}
\usepackage{amsmath,amssymb,amsfonts}
\usepackage{algorithmic}
\usepackage{graphicx}
\usepackage{textcomp}
\usepackage{wrapfig,colortbl}
\usepackage{hyperref}
\hypersetup{colorlinks=true, urlcolor=nblue}
\usepackage{textgreek}
\usepackage[inkscapelatex=false]{svg} 
\usepackage[final]{changes}
\usepackage{multirow} 
\usepackage{subfiles} 
\usepackage{lipsum}
\definecolor{abstractbg}{rgb}{1,0.969,0.914}
\def\BibTeX{{\rm B\kern-.05em{\sc i\kern-.025em b}\kern-.08em
    T\kern-.1667em\lower.7ex\hbox{E}\kern-.125emX}}
\begin{document}
\title{Thermal Endurance of Suspended Thin-Film Lithium Niobate up to 800 $^\circ$C }
\author{Mihir Chaudhari, Lezli Matto, Naveed Ahmed, Michael Liao, Vivek Tallavajhula, Yidou Long, Ziqian Yao, Joshua Campbell, Tzu-Hsuan Hsu, Mark S. Goorsky, and Ruochen Lu 
\thanks{This paper is an expanded version of the International Conference on Solid-State Sensors, Actuators and Microsystems (Transducers) 2025.}
\thanks{The paper was submitted on July 30, 2025. This work was supported by the Defense Advanced Research Projects Agency (DARPA) High Operational Temperature Sensors (HOTS) program.}
\thanks{Mihir Chaudhari, Naveed Ahmed, Vivek Tallavajhula, Ziqian Yao, Joshua Campbell, Tzu-Hsuan Hsu, and Ruochen Lu are with the Chandra Family Department of Electrical and Computer Engineering, The University of Texas at Austin, Austin, TX 78712 USA (e-mail: mihirc@utexas.edu). Lezli Matto, Michael Liao, Yidou Long, and Mark S. Goorsky are with the Department of Materials Science and Engineering, University of California, Los Angeles, Los Angeles, CA 90095 USA.}}

\IEEEtitleabstractindextext{%
\fcolorbox{abstractbg}{abstractbg}{%
\begin{minipage}{\textwidth}\rightskip2em\leftskip\rightskip\bigskip
\begin{wrapfigure}[24]{r}{3in}%
\hspace{-3pc}\includesvg[width=2.9in]{graphical_abstract.svg}
\end{wrapfigure}%
\begin{abstract}
The need for high-temperature piezoelectric microelectromechanical systems (MEMS) requires pushing piezoelectric platforms to their thermal limits. In harsh thermal environments, piezoelectric MEMS devices are expected to sustain severe damage because of material degradation and coefficient of thermal expansion (CTE) mismatches between the functional layers and the carrier wafer. This paper investigates the thermal endurance of the suspended thin-film lithium niobate (LN) platform by observing the structural integrity and performance of acoustic Lamb wave resonators after annealing rounds at increasing temperatures, with a focus on temperatures between 550 $^\circ$C and 800 $^\circ$C, with 50 $^\circ$C temperature increments. Fundamental symmetric (S0) mode acoustic resonators are fabricated on 600 nm stoichiometric LN (sLN) with 40 nm thick platinum top electrodes and a thin titanium adhesion layer. After each annealing round, changes in the devices' resonant frequency and quality factor (\emph{Q}) are quantitatively studied. The devices and material stack are further analyzed with resistivity structures, optical microscope images, and X-ray diffraction (XRD) measurements. The results provide valuable insights into the design and material selection necessary to optimize the suspended thin-film LN platform for high temperatures. Understanding the thermal limit of the platform enables its use for sensors, actuators, resonators, and potentially other thin-film LN microsystems, e.g, photonics, electro-optical, and acousto-optical systems in harsh thermal environments.
\end{abstract}

\bigskip
\end{minipage}}}

\maketitle

\section{Introduction}
\label{sec:introduction}

\IEEEPARstart{H}{igh}-temperature piezoelectric microelectromechanical systems (MEMS) are increasingly attractive for applications such as power plants, petroleum reactors, automotive combustion engines, and aerospace systems, where devices must operate reliably at temperatures exceeding 700~$^\circ$C \cite{transducers, background1, background2, background3}. MEMS enable compact, low-power, and high-sensitivity sensors, actuators, and resonators; however, conventional MEMS technologies—particularly those based on silicon—are typically limited to operating temperatures below 600~$^\circ$C due to intrinsic material limitations \cite{sisensor1,sisensor2}. Integrating MEMS and fiber-optic systems can result in devices with operating temperatures exceeding 1000 $^\circ$C, but they require more bulky and complex setups that are not suitable for miniaturized or integrated sensing applications \cite{background3} \cite{opticsensor1}. To reach higher device operating temperatures while retaining the benefits of a pure MEMS-based approach, alternative piezoelectric platforms, such as lithium niobate (LN), show promise for operating in harsh thermal environments \cite{materialln1, materialln2, materialln3}.

\begin{table*}[!t]
\arrayrulecolor{subsectioncolor}
\setlength{\arrayrulewidth}{1pt}
{\sffamily\bfseries\begin{tabular}{lp{6.75in}}\hline
\rowcolor{abstractbg}\multicolumn{2}{l}{\color{subsectioncolor}{\itshape
Highlights}{\Huge\strut}}\\
\rowcolor{abstractbg}$\bullet$ & \deleted{Demonstrates} \added{Studies} thermal endurance of suspended thin-film lithium niobate (LN) resonators up to \deleted{750} \added{800} $^\circ$C.\\
\rowcolor{abstractbg}$\bullet${\large\strut} & Presents comprehensive device- and material-level analysis of thermal degradation mechanisms in suspended Pt electrodes on LN platforms.\\
\rowcolor{abstractbg}$\bullet${\large\strut} & Provides design guidelines for high-temperature thin-film LN platforms, targeting MEMS and other photonics, electro-optical, and acousto-optical applications.\\[2em]\hline
\end{tabular}}
\setlength{\arrayrulewidth}{0.4pt}
\arrayrulecolor{black}
\end{table*}

High-temperature operation is being studied on commonly used piezoelectric platforms including lead zirconate titanate (PZT) \cite{pzt1, pzt2, pzt3}, aluminum nitride (AlN) \cite{aln1, aln2, aln3}, scandium aluminum nitride (AlScN) \cite{alscn1, alscn2, alscn3}, and LN \cite{ln1, ln2, ln3}. Table I, which is adapted from \cite{platformcompare}, provides a comparison of these platforms. \added{A comparison of high temperature studies on these platforms is provided in Table II.} Among these platforms, LN is desirable due to its higher figures of merit, which feature\added{s} an exceptionally large electromechanical coupling coefficient. Additionally, LN possesses a \deleted{high} Curie temperature of 1200 $^\circ$C, which has potential for use in harsh thermal environments. The thermal endurance of LN is also affected by its composition. Specifically, stoichiometric LN (sLN) is reported to be more effective for high-temperature operation compared to congruent LN (cLN). sLN has a near 1:1 ratio of lithium and niobium, while cLN has a lower concentration of lithium compared to niobium \cite{materialln1}. cLN has lithium vacancy defects, and the material degrades above 300 $^\circ$C, while sLN is stable until at least 900 $^\circ$C \cite{materialln1}. The LN quality can degrade when heated because lithium atoms can diffuse out of the LN, and can form Li-poor LiNb\textsubscript{3}O\textsubscript{8} phases in the LN film \cite{materialln2}. Although work on LN for high temperatures exists, most studies focus on surface acoustic wave (SAW)-based devices on solidly mounted platforms \cite{ln4, ln5, ln6, ln7, ln8, ln9}. The thermal survivability of the suspended thin-film LN platform has not been extensively studied until the platform's temperature limit.

\begin{table}
\caption{Piezoelectric MEMS Platforms Comparison}
\label{tab:platformcompare}
\setlength{\tabcolsep}{3pt}
\begin{tabular}{
  >{\centering\arraybackslash}m{36pt}
  >{\centering\arraybackslash}m{15pt}
  >{\centering\arraybackslash}m{15pt}
  >{\centering\arraybackslash}m{15pt}
  >{\centering\arraybackslash}m{26pt}
  >{\centering\arraybackslash}m{22pt}
  >{\centering\arraybackslash}m{22pt}
  >{\centering\arraybackslash}m{53pt}
}
\hline\hline
\multirow{2}{36pt}{Material} & \multirow{2}{15pt}{Curie T ($^\circ$C)} & \multirow{2}{15pt}{E-Field} & \multirow{2}{15pt}{Mode} & Resonator FoM & \multicolumn{3}{c}{Sensor/Transducer FoMs} \\
& & & & $e^2/(\varepsilon c)$ & $e$ (C/m\textsuperscript{2}) & $e/\varepsilon$ (GV/m) & $e/(\varepsilon \tan \delta)^{1/2}$ (10\textsuperscript{6} (J/m\textsuperscript{3})\textsuperscript{1/2}) \\
\hline
PZT 5A & 350 & TFE & LE & 0.3\% & 18.7 & 1.30 & 1.1 \\
AlN & 1150 & TFE & LE & \replaced{0.97}{0.96}\% & 0.59 & \replaced{6.50}{0.65} & \replaced{1.96}{0.62} \\
AlN & 1150 & TFE & TE & \replaced{6.36}{6.0}\% & 1.47 & \replaced{16.12}{1.61} & \replaced{4.87}{1.54} \\
Al\textsubscript{0.7}Sc\textsubscript{0.3}N & 1100 & TFE & LE & 0.84\% & 0.70 & 3.75 & \replaced{0.85}{2.70} \\
Al\textsubscript{0.7}Sc\textsubscript{0.3}N & 1100 & TFE & TE & \replaced{12.90}{9.7}\% & 2.38 & \replaced{12.74}{1.27} & \replaced{2.90}{0.92} \\
\textbf{X-cut LiNbO\textsubscript{3}} & \textbf{1200} & \textbf{LFE} & \textbf{LE} & 3\textbf{1.8\%} & \textbf{4.61} & \textbf{13.24} & \textbf{6.85} \\
\hline\hline
\end{tabular}
\end{table}

The unique characteristics of the suspended platform support bulk acoustic waves (BAWs) and Lamb waves. Lamb waves form in thin plates with parallel free boundaries and tend to have greater electromechanical coupling than their SAW counterparts. The cause for the larger coupling can be intuitively thought of as the electrical and mechanical fields being confined in a more compact area, thereby enhancing cross-domain interaction. A challenge of the suspended platform is that it results in more fragile structures that are prone to collapsing, and they tend to break in harsh conditions. The metal, LN, and support substrate all have different coefficients of thermal expansion (CTEs) and are expected to crack when heated to and cooled from high temperatures. Previous works have tested suspended LN devices up to temperatures of 550 $^\circ$C and 500 $^\circ$C, and demonstrate the devices' operability at these temperatures \cite{transducers} \cite{susln}. However, the platform has not been pushed to its thermal limit, and its high-temperature failure mechanisms have not been studied.

\begin{table}[!t]
\centering
\caption{High Temperature Piezoelectric Studies}
\label{tab:metalcompare}
\setlength{\tabcolsep}{3pt}
\begin{tabular}{
  >{\centering\arraybackslash}m{40pt}
  >{\centering\arraybackslash}m{50pt}
  >{\centering\arraybackslash}m{50pt}
  >{\centering\arraybackslash}m{40pt}
  >{\centering\arraybackslash}m{30pt}
}
\hline\hline
Platform & Highest Temperature Studied ($^\circ$C) & Device Studied & Heating method & Reference \\
\hline
PZT & 550 & High-overtone bulk acoustic resonator & Heated stage & [11] \\
AlN & 700 & Lamb wave resonator & Infrared lamp & [14] \\
Al\textsubscript{0.7}Sc\textsubscript{0.3}N & 1000 & Capacitors & Custom vacuum chamber & [19] \\
Solidly mounted LiNbO\textsubscript{3} & 800 & Surface acoustic wave temperature sensor & Muffle furnace & [24] \\
\textbf{Suspended LiNbO\textsubscript{3}} & \textbf{800} & \textbf{Lamb wave resonator} & \textbf{Anneal oven} & \textbf{This work} \\
\hline\hline
\end{tabular}
\end{table}

In this work, we study thin-film suspended LN acoustic resonators after they are subjected to increasingly higher anneal temperatures, until the devices are inoperable. Between annealing rounds, the resonators are characterized using optical microscope images to examine their physical structure, electrode metal resistivity to determine the metal quality, radio frequency (RF) electrical measurements to assess the device's performance, and X-ray diffraction (XRD) measurements to further evaluate the material quality. These characterizations help determine failure points of the acoustic resonators and the suspended platform. Understanding the thermal limit of the suspended platform will not only provide insight into designing MEMS devices that are more resilient to harsh thermal environments but also enable more suspended thin-film LN systems for higher-temperature applications, such as photonic, electro-optic, and acousto-optic systems.

This work is an extension of \cite{transducers}, which focused on the thermal survivability of the resonators up to 550 $^\circ$C. The focus of this work is on the annealing temperature range of 550 $^\circ$C to 800 $^\circ$C, where significant damage occurs to the devices as the LN delaminates and severely cracks along one of its crystal axes, and the metal loses conductivity. These damages lead to the failure of the acoustic resonators. The results from the annealing rounds in this work, which study the devices until their failure, provide valuable insights into how to better design for high-temperature operation on the suspended LN platform. This paper is organized as follows. Section II outlines the design considerations required to create high-temperature suspended LN acoustic resonators, as well as the fabrication process for these devices. Section III summarizes the results from \cite{transducers}, focusing on the results until 550 $^\circ$C. Section IV focuses on new results that test the devices until their failure at annealing temperatures ranging from 550 $^\circ$C to 800 $^\circ$C. The study ends after the 800 $^\circ$C anneal, when a resonator of interest is no longer operable.

\begin{table}[!t]
\centering
\caption{Comparison of Metals for MEMS Applications}
\label{tab:metalcompare}
\setlength{\tabcolsep}{3pt}
\begin{tabular}{
  >{\centering\arraybackslash}m{70pt}
  >{\centering\arraybackslash}m{70pt}
  >{\centering\arraybackslash}m{70pt}
}
\hline\hline
Metal&
Melting point ($^\circ$C)&
$\rho$ ($\mu \Omega \cdot cm$) \\
\hline
Aluminum & 660 & 2.65 \\
Silver & 961 & 1.59 \\
Gold & 1063 & 2.24 \\
Copper & 1084 & 1.72 \\
Titanium & 1670 & 43 \\
Chromium & 1860 & 13 \\
Tungsten & 3400 & 5.65 \\
\textbf{Platinum} & \textbf{1770} & \textbf{10.5} \\
\hline\hline
\end{tabular}
\end{table}

\begin{figure}[!t]
\centerline{\includesvg[width=\columnwidth]{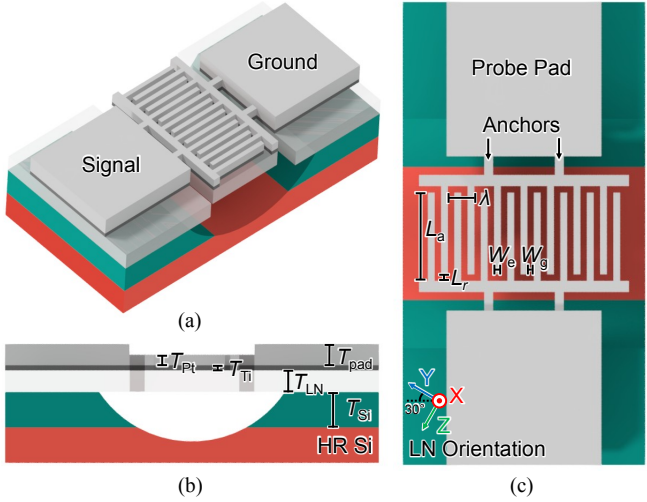}}
\caption{Schematic of an example suspended LN resonator designed for high-temperature survival with (a) a side view, (b) a cross-section view, and (c) a top view. Key dimensions are given in Table IV.}
\label{fig1}
\end{figure}

\section{Design, Simulation, and Fabrication}

The material selection serves as the foundation for high-temperature devices, and the two main materials to consider for piezoelectric acoustic resonators are the piezoelectric layer and the metal used for electrodes. Thin-film LN is anisotropic and can be cut from different crystal orientations. X-cut (where the X axis is perpendicular to the plane of the film) was selected because of its versatile usage in various LN systems, including MEMS and optical systems \cite{xcutapp1, xcutapp2, xcutapp3, xcutapp4, xcutapp5}. The LN is transferred from a bulk wafer and bonded onto the Si support substrate. The metal is selected based on its melting point and resistivity, and a comparison of commonly used MEMS metals is provided in Table III \cite{metaltable1, metaltable2}. Platinum was selected as the electrode metal because of its high melting point. Although tungsten exhibits a higher melting point and lower resistivity, it is etched in the later Si isotropic etch process. At high temperatures, platinum films gain defects such as holes, grain coarsening, and hillocks; however, an adhesion layer can make the Pt more resilient to defects, and improve adhesion to the LN beneath it \cite{pt1, pt2, pt3}. Titanium was used as the adhesion layer. While Cr is another commonly used adhesion layer, Ti offers better stability after high temperature annealing with platinum \cite{tibetter}.

The resonators are designed for 600 nm X-cut sLN on 1 \textmu{}m amorphous Si (a-Si) on 500 \textmu{}m high-resistivity silicon (HR Si), with 40 nm of Pt for the electrodes and 5 nm of Ti for the adhesion layer, and target the fundamental symmetric (S0) mode as the main acoustic mode. The design and key dimensions of one example resonator are given in Fig. 1 and Table IV. There is a 30$^\circ$ in-plane rotation to maximize the electromechanical coupling \cite{lwr}. Additional resonators are designed by varying the in-plane rotation, the anchor design (no anchors, one anchor, or two anchors), and the number and width of the interdigitated electrodes. \added{Sixty resonators are designed in total.} A variety of commonly used resonator designs \replaced{are}{is} included to prevent the thermal endurance study from being dominated by a single design. The anchor design is a trade-off between structural integrity and quality factor (\emph{Q}). With the exception of no anchors, fewer anchors will allow the resonator to vibrate more freely, but will leave the resonator more prone to collapsing. The design of the interdigitated electrodes affects the static capacitance, \emph{Q}, power handling, and resonant frequency. The pitch of the interdigitated fingers determines the acoustic wavelength, which in turn sets the frequency of the resonator. Alongside the resonators, meandering line resistivity structures (Fig. 2) are designed with the same metal stack as the resonators' electrodes to monitor the metal quality through metal resistivity. The metal resistivity directly affects the resonators' performance, as increased resistivity leads to more electrical energy loss and lowers the \emph{Q} of the device.

\begin{table}[!t]
\caption{Example Resonator Key Dimensions}
\label{tab:resonatordimensions}
\setlength{\tabcolsep}{3pt}
\begin{tabular}{
  >{\centering\arraybackslash}m{20pt}
  >{\centering\arraybackslash}m{45pt}
  >{\centering\arraybackslash}m{25pt}
  >{\centering\arraybackslash}m{20pt}
  >{\centering\arraybackslash}m{80pt}
  >{\centering\arraybackslash}m{25pt}
}
\hline\hline
Sym. & Parameter & Value & Sym. & Parameter & Value \\
\hline
$L_a$ & Aperture & 43 $\mu$m & $T_{Pt}$ & Electrode thickness & 40 nm \\
$L_r$ & Gap length & 3 $\mu$m & $T_{Ti}$ & Adhesion layer thickness & 5 nm \\
$\lambda$ & Wavelength & 12 $\mu$m & $T_{pad}$ & Probe pad thickness & 80 nm \\
$W_e$ & Electrode width & 3 $\mu$m & $T_{LN}$ & Piezoelectric thickness & 600 nm \\
$W_g$ & Gap width & 3 $\mu$m & $T_{Si}$ & Sacrificial layer thickness & 1 $\mu$m \\
\hline\hline
\end{tabular}
\end{table}

\begin{figure}[!t]
\centerline{\includesvg[width=\columnwidth]{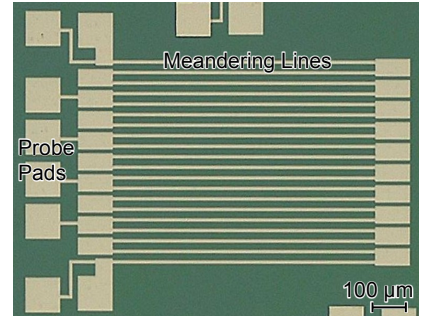}}
\caption{Meandering line resistivity testing structure. It has the same metal stack as the resonators to monitor the metal quality through resistivity.}
\label{fig2}
\end{figure}

The resonator design is verified with COMSOL finite element analysis (FEA). The 3D FEA model has a pair of interdigitated electrodes on top of LN on top of an air layer. Free boundaries are applied to the electrodes and the LN. The resonant modes are identified with an eigenfrequency analysis, and three modes were selected for the study: the S0 mode, a higher-order fundamental shear horizontal (SH0) mode, and a higher-order S0 mode. The electrical admittance is obtained by driving the electrodes with an AC voltage and a frequency domain sweep. The simulation results for the acoustic mode shapes and the electrical admittance are shown in Fig. 3.

\begin{figure}[!t]
\centerline{\includesvg[width=\columnwidth]{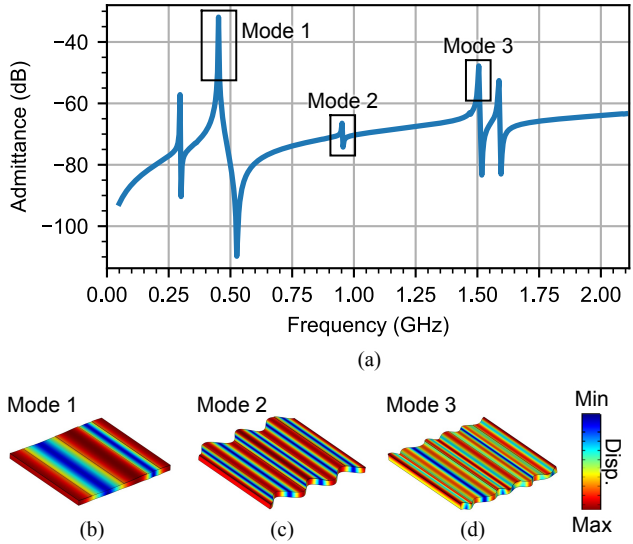}}
\caption{(a) Simulated example resonator admittance with (b)-(d) displacements of the modes of interest. (b) Mode 1 is the S0 mode, (c) mode 2 is a higher-order fundamental SH0 mode, and (d) mode 3 is a higher-order S0 mode.}
\label{fig3}
\end{figure}

The resonators and resistivity devices are fabricated in-house on the same substrate. A schematic of the resonator fabrication process is shown in Fig. 4. First, release windows are etched through the LN and sacrificial amorphous silicon (a-Si) layers using argon gas ion milling. Next, metal is deposited onto the LN by electron beam evaporation in two deposition rounds. In the first round, 5 nm of Ti is followed by 40 nm of Pt, which is deposited onto the LN for the electrodes. In the second round, an additional 40 nm of Pt is deposited to thicken the probe pads. No metal from the second round is deposited onto the meandering line resistivity structures, so that they have the same thickness as the resonator electrodes. After each metal deposition round, the unwanted metal is removed with a lift-off process. Finally, the devices are released, or suspended, by a XeF\textsubscript{2} etch.

\begin{figure}[!t]
\centerline{\includesvg[width=\columnwidth]{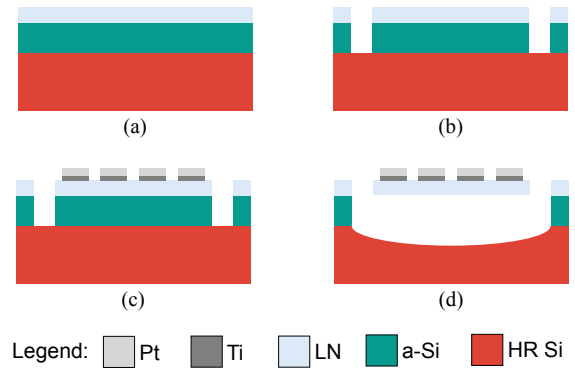}}
\caption{Fabrication flow. (a) The initial stack is etched to create (b) release windows for device suspension. Next, the (c) Ti adhesion layer and Pt electrodes are deposited. Finally, (d) the device is suspended with a XeF\textsubscript{2} etch.}
\label{fig4}
\end{figure}

\begin{figure}[!t]
\centerline{\includesvg[width=\columnwidth]{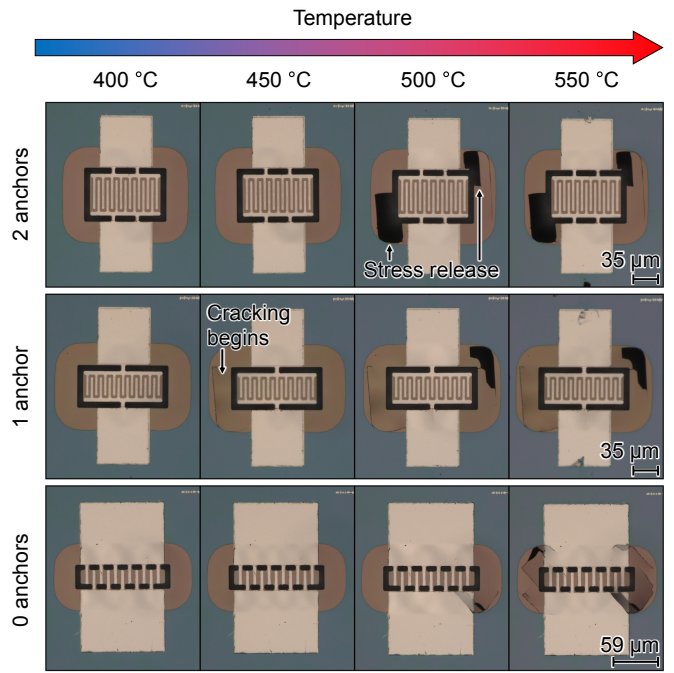}}
\caption{Optical images of \deleted{example} resonators after annealing between 400~$^\circ$C and 550~$^\circ$C. \added{The 2-anchor resonator is the example resonator.} In this temperature range, cracks in the LN are visible in the microscope.}
\label{fig5}
\end{figure}

The performance of 60 resonators is tracked throughout the study. The fabricated example resonator \added{which is the design presented in Fig. 1,} has a similar pre-anneal response compared to its simulation. It contains the three modes of interest, as well as spurious modes. The presence of the main modes is sufficient for studying thermal endurance, as resonant parameters can be extracted and compared across different annealing rounds. For the example resonator, it has mode 1 with a resonant frequency of 467.8 MHz and a \emph{Q} of 458, mode 2 with a resonant frequency of 958.1 MHz and a \emph{Q} of 1398, and mode 3 with a resonant frequency of 1.475 GHz and a \emph{Q} of 717.

\section{Thermal Endurance up to 550 $^\circ$C}

To test the thermal endurance of the suspended thin-film LN platform, we performed a cycle of characterization and annealing. The devices are characterized at room temperature in air, with a focus on the thermal effects following an annealing round. The anneal is performed in vacuum, with a ramp rate of 100 $^\circ$C\added{/hour} for heating/cooling to/from the target temperature. The ramp rate is selected to avoid the pyroelectric effects of LN \cite{thermalshock, pyroelectric}. The target temperature is maintained at a constant level for 10 hours. The first anneal has a target temperature of 250 $^\circ$C, and subsequent rounds increase the target temperature by 50 $^\circ$C.

\begin{table}[t]
\centering
\caption{Metal Resistivity after Annealing}
\label{tab:resistivity550}
\setlength{\tabcolsep}{3pt}
\begin{tabular}{
  >{\centering\arraybackslash}m{65pt}
  >{\centering\arraybackslash}m{35pt}
  >{\centering\arraybackslash}m{45pt}
  >{\centering\arraybackslash}m{60pt}
}
\hline\hline
Annealing temperature ($^\circ$C) & $\rho$ ($\mu \Omega \cdot cm$) & Fractional $\Delta \rho$ (\%) & Normalized to Bulk Pt $\rho$ \\
\hline
Initial & 35.39 & 0 & 3.37 \\
250 & 29.59 & -16.39 & 2.44 \\
300 & 30.90 & -12.69 & 2.94 \\
350 & 32.46 & -8.28 & 3.09 \\
400 & 35.05 & -0.96 & 3.34 \\
450 & 38.80 & 9.64 & 3.70 \\
500 & 42.79 & 20.91 & 4.08 \\
550 & 48.20 & 36.20 & 4.59 \\
\hline\hline

\end{tabular}
\end{table}

\begin{figure}[!t]
\centerline{\includesvg[width=\columnwidth]{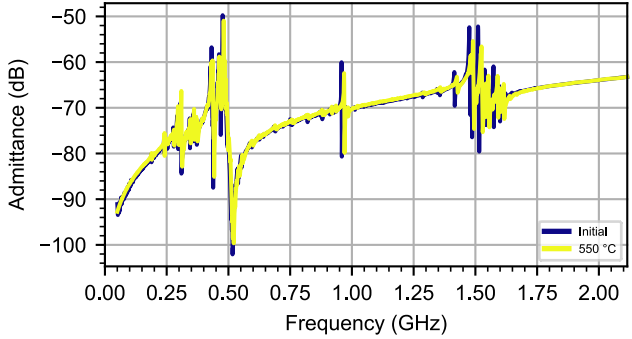}}
\caption{Measured wideband admittance for the example resonator before the initial anneal and after the 550 $^\circ$C anneal, showing modes 1 to 3 as in the simulation.}
\label{fig6}
\end{figure}

The optical microscope images reveal minor visual changes in the physical structure until approximately 500 $^\circ$C (Fig. 5). For the devices shown, no visual changes are observed before 400 $^\circ$C. After 500 $^\circ$C, the LN cracks due to stress release, as indicated by the black spots next to the devices. Although the cracking occurs in the suspended region, it does not cause the devices to collapse. The devices may crack after a lower temperature annealing round, but will mostly maintain their appearance after cracking until 550 $^\circ$C.

The resistivity drops after the first anneal at 250 $^\circ$C, and increases with subsequent annealing rounds (Table V). The resistivity is calculated from averaged resistance measurements with DC probes on the meandering line resistivity structures. After the initial anneal, the resistivity drops significantly by 16.39\%. The initial anneal can clean defects from the metal film, but subsequent anneals increase the resistivity. After an anneal at 450 $^\circ$C the resistivity is larger than its initial value, having increased to 38.80 \textmu{}$\Omega\cdot$cm from 35.39 \textmu{}$\Omega\cdot$cm. After 550 $^\circ$C, the resistivity increases to 48.20 \textmu{}$\Omega\cdot$cm. Although no changes are visible from the optical microscope images, defects such as grain coarsening are forming in the metal structure \cite{pt1}.

\begin{figure}[!t]
\centerline{\includesvg[width=\columnwidth]{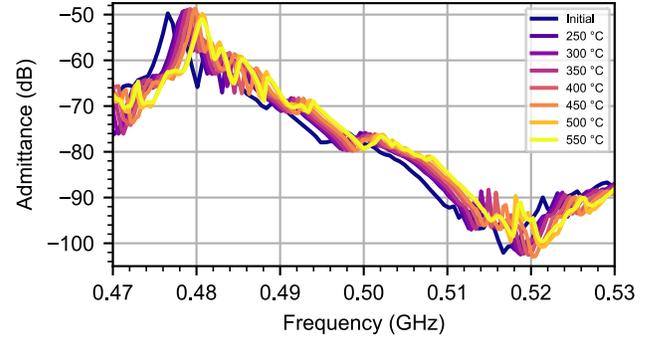}}
\caption{Measured admittance of the example resonator at mode 1 before and after annealing rounds ranging from 250 $^\circ$C to 550 $^\circ$C.}
\label{fig7}
\end{figure}

\begin{table}[!t]
\centering
\caption{Mode 1 Resonance Parameters}
\label{tab:mode1at550}
\setlength{\tabcolsep}{3pt}
\begin{tabular}{>{\centering\arraybackslash}m{65pt}>{\centering\arraybackslash}m{50pt}>{\centering\arraybackslash}m{30pt}>{\centering\arraybackslash}m{30pt}>{\centering\arraybackslash}m{40pt}}
\hline\hline
Annealing temperature ($^\circ$C) & Resonant $f$ (MHz) & Fractional $\Delta f$ (\%) & $Q$ & Fractional $\Delta Q$ (\%) \\
\hline
Initial & 467.8 & 0 & 458 & 0 \\
250 & 478.3 & 2.24 & 431 & -5.96 \\
300 & 478.5 & 2.29 & 435 & -4.98 \\
350 & 478.9 & 2.37 & 450 & -1.81 \\
400 & 479.4 & 2.48 & 453 & -1.18 \\
450 & 479.5 & 2.50 & 454 & -0.98 \\
500 & 480.7 & 2.76 & 348 & -23.97 \\
550 & 480.7 & 2.76 & 330 & -28.03 \\
\hline\hline
\end{tabular}
\end{table}

\begin{figure}[!t]
\centerline{\includesvg[width=\columnwidth]{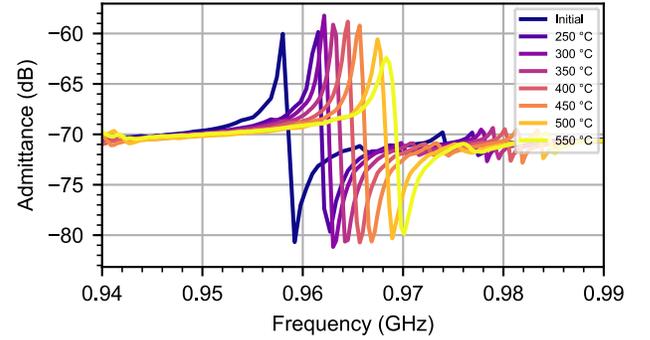}}
\caption{Measured admittance of the example resonator at mode 2 before and after annealing rounds ranging from 250 $^\circ$C to 550 $^\circ$C.}
\label{fig8}
\end{figure}

\begin{table}[!t]
\centering
\caption{Mode 2 Resonance Parameters}
\label{tab:mode2at550}
\setlength{\tabcolsep}{3pt}
\begin{tabular}{>{\centering\arraybackslash}m{65pt}>{\centering\arraybackslash}m{50pt}>{\centering\arraybackslash}m{30pt}>{\centering\arraybackslash}m{30pt}>{\centering\arraybackslash}m{40pt}}
\hline\hline
Annealing temperature ($^\circ$C) & Resonant $f$ (MHz) & Fractional $\Delta f$ (\%) & $Q$ & Fractional $\Delta Q$ (\%) \\
\hline
Initial & 958.1 & 0 & 1398 & 0 \\
250 & 961.4 & 0.34 & 1529 & 9.32 \\
300 & 962.2 & 0.43 & 2008 & 43.62 \\
350 & 963.2 & 0.53 & 1885 & 34.78 \\
400 & 964.5 & 0.67 & 1850 & 32.31 \\
450 & 965.7 & 0.79 & 1679 & 20.03 \\
500 & 967.7 & 1.00 & 1233 & -11.85 \\
550 & 968.7 & 1.11 & 851 & -39.13 \\
\hline\hline
\end{tabular}
\end{table}

The resonators experience a frequency upshift and have different trends for their \emph{Q} depending on the acoustic mode. The resonators are measured with two-port ground-source-ground (GSG) probes and a Keysight vector network analyzer (VNA) in air at room temperature after an anneal round. The wideband admittance for the example resonator is shown in Fig. 6, which displays the modes of interest observed in the simulation, as well as resonant frequency upshifts. The admittance for each mode is demonstrated individually in Figs. 7-9. The extracted resonant frequencies and quality factors \emph{Q}s, along with their fractional change from their initial values, are reported in Tables VI-VIII. The resonance characteristics are extracted by fitting a modified Butterworth Van-Dyke (MBVD) model to the admittances. In mode 1, the \emph{Q} remains below its original value before annealing. In modes 2 and 3, the \emph{Q} rises above the initial value until after 500 $^\circ$C and 550 $^\circ$C anneals, respectively, which are both beyond the point where the metal resistivity increases above its original value. \deleted{Considering the variation in \emph{Q} trends, there is no direct correlation between the electrode resistivity and the device \emph{Q}.} The targeted S0 mode does not experience a \emph{Q} enhancement, while the higher-order modes do experience a \emph{Q} enhancement after certain annealing temperatures, likely due to more vibrations in the electrodes. More work will be required to investigate the specific changes, as they fall outside the scope of this work on the thermal endurance of suspended LN resonators. Overall, the resonators demonstrate a strong thermal endurance up to 550 $^\circ$C.

\section{Analysis from 550 $^\circ$C to 800 $^\circ$C}

After annealing rounds at temperatures ranging from 550 $^\circ$C to 750 $^\circ$C, the devices and material stack sustain significant damage: resonator anchor breakage, LN delamination, new cracks in the entire LN layer, and hole formation in the metal. Fig. 10 illustrates the changes in the resonators' appearance within this annealing range, and Fig. 11 focuses on the new defects observed after the 700 $^\circ$C anneal. In the 750 $^\circ$C and 800 $^\circ$C annealing rounds, the annealing time is reduced to 1 hour from 10 hours as the annealing furnace's operating limits are reached. After the 800 $^\circ$C anneal, the resonator\added{s} become\deleted{s} inoperable due to metal degradation and anchor breakage.

\begin{figure}[!t]
\centerline{\includesvg[width=\columnwidth]{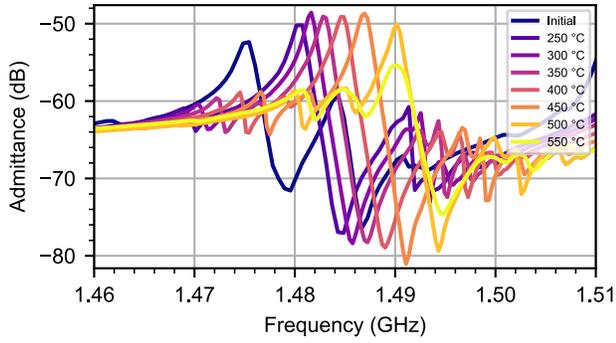}}
\caption{Measured admittance of the example resonator at mode 3 before and after annealing rounds ranging from 250 $^\circ$C to 550 $^\circ$Cs.}
\label{fig9}
\end{figure}

\begin{table}[!t]
\centering
\caption{Mode 3 Resonance Parameters}
\label{tab:mode3at550}
\setlength{\tabcolsep}{3pt}
\begin{tabular}{>{\centering\arraybackslash}m{65pt}>{\centering\arraybackslash}m{50pt}>{\centering\arraybackslash}m{30pt}>{\centering\arraybackslash}m{30pt}>{\centering\arraybackslash}m{40pt}}
\hline\hline
Annealing temperature ($^\circ$C) & Resonant $f$ (MHz) & Fractional $\Delta f$ (\%) & $Q$ & Fractional $\Delta Q$ (\%) \\
\hline
Initial & 1.475 & 0 & 717 & 0 \\
250 & 1.481 & 0.41 & 970 & 35.25 \\
300 & 1.482 & 0.47 & 1163 & 62.15 \\
350 & 1.483 & 0.54 & 1096 & 52.91 \\
400 & 1.485 & 0.68 & 1072 & 49.57 \\
450 & 1.487 & 0.81 & 1116 & 55.62 \\
500 & 1.490 & 1.02 & 1059 & 47.68 \\
550 & 1.490 & 1.02 & 631 & -11.98 \\
\hline\hline
\end{tabular}
\end{table}

\subsection{Material Quality}

\replaced{Significant physical damage is visible after the 700 $^\circ$C anneal}{There is significant physical damage visible after the 700 $^\circ$C anneal.} After the 600 $^\circ$C anneal, the LN cracks more. After the 650 $^\circ$C anneal, the resonators begin bending as seen by the partially defocused electrodes in the microscope image. After the 700 $^\circ$C anneal, the LN cracks along its Z axis. The LN delaminates and creates ring patterns extending in the LN's Y axis, originating from the Si cracks.

\begin{figure}[!t]
\centerline{\includesvg[width=\columnwidth]{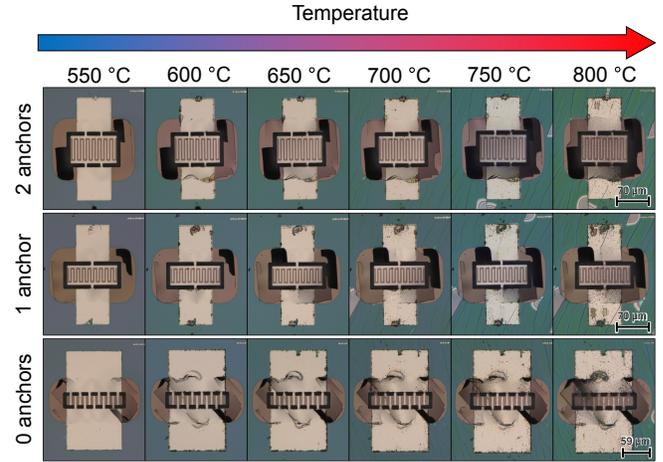}}
\caption{Optical microscope images of resonators after annealing rounds ranging from 550 $^\circ$C to 750 $^\circ$C. The device with one anchor shown here fails after the 650 $^\circ$C anneal.}
\label{fig10}
\end{figure}

\begin{figure}[!t]
\centerline{\includesvg[width=\columnwidth]{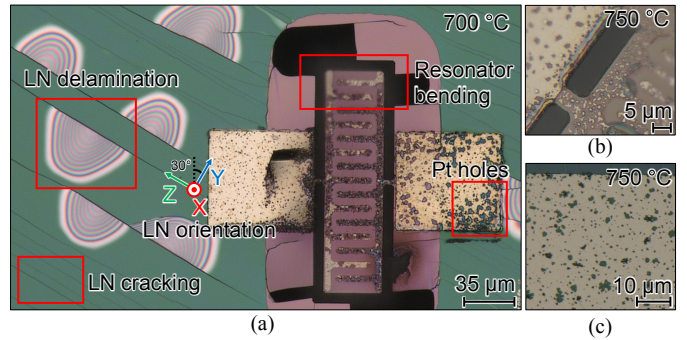}}
\caption{(a) Damages seen after the 700 $^\circ$C anneal. (b) and (c) provide close-ups of Pt holes in a resonator and resistivity structure respectively.}
\label{fig11}
\end{figure}

\begin{figure}[!t]
\centerline{\includesvg[width=\columnwidth]{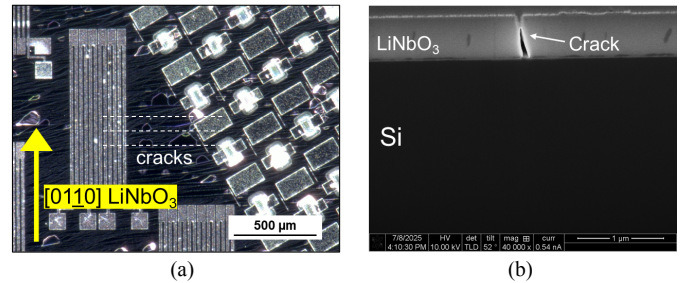}}
\caption{(a) Optical microscopy of the sample after cracking, where cracks perpendicular to the [01\underline{1}0] LN direction are observed, and (b) FIB cross-section showing cracking occurs on the LN film and does not propagate through the Si.}
\label{fig12}
\end{figure}

An optical microscopy image (Fig. 12(a)) shows the orientation of the cracks observed after the 700 $^\circ$C annealing step. Cracks are aligned perpendicular to the [01\underline{1}0] LN direction. This is the largest in-plane coefficient of thermal expansion (CTE) direction for X-cut LN \deleted{with a value of} \added{(}14.4×10\textsuperscript{-6} $^\circ$C\textsuperscript{-1}\added{)} \cite{ucla1}, which is much higher than that of Si with a value of 2.6×10\textsuperscript{-6} $^\circ$C\textsuperscript{-1} at room temperature \cite{ucla2}. Additionally, the (01\underline{1}0) LN plane corresponds to a secondary low-energy cleavage plane \cite{ucla3}. The combination of these two factors is thought to be the reason why cracks form preferentially perpendicular to the [01\underline{1}0] LN direction after annealing at 700 $^\circ$C. A cross-sectional focused ion beam (FIB)/scanning electron microscope (SEM) image (Fig. 12(b)), generated with a FEI Nova 600 DualBeam, shows the crack forms in the LN film and does not propagate through the Si.

\begin{figure}[!t]
\centerline{\includesvg[width=\columnwidth]{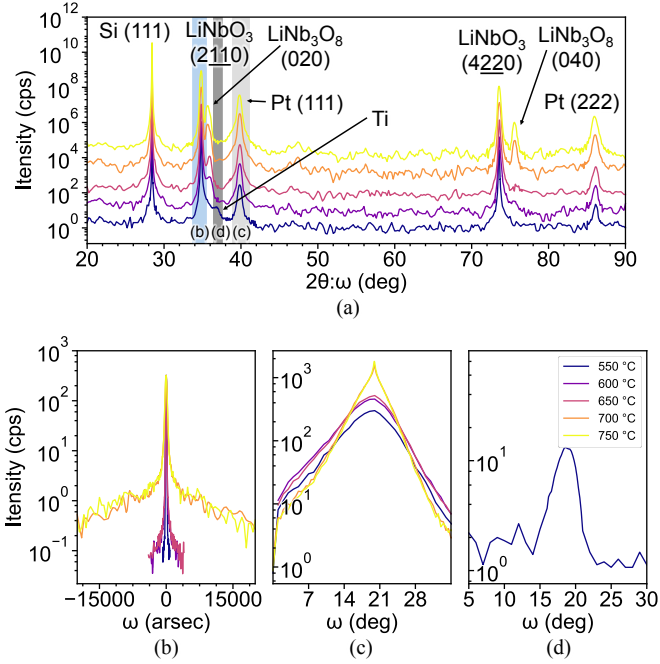}}
\caption{(a) DAD  2\straighttheta:\textomega\ scan of the stack after each annealing cycle starting at 550 $^\circ$C. (b) TAD Rocking curve of the symmetric (2\underline{11}0) LN. DAD rocking curve of the (c) Pt film and (d) Ti film.}
\label{fig13}
\end{figure}

\begin{table}
\caption{XRD Peak Widths}
\label{tab:platformcompare}
\setlength{\tabcolsep}{3pt}
\begin{tabular}{
  >{\centering\arraybackslash}m{55pt}
  >{\centering\arraybackslash}m{22pt}
  >{\centering\arraybackslash}m{34pt}
  >{\centering\arraybackslash}m{46pt}
  >{\centering\arraybackslash}m{21pt}
  >{\centering\arraybackslash}m{37pt}
}
\hline\hline
\multirow{2}{60pt}{\centering Annealing temperature ($^\circ$C)} & \multicolumn{3}{c}{LiNbO\textsubscript{3} (2\underline{11}0)} & \multicolumn{2}{c}{Pt (111)} \\
& FWHM & FW(0.01)M & FW(0.001)M & FWHM & FW(0.01)M \\
\hline
550 & \phantom{$^\circ$}172$^\circ$ & \phantom{$^\circ$}460$^\circ$ & \phantom{$^\circ$}570$^\circ$ & \phantom{$^\circ$}9$^\circ$ & \phantom{$^\circ$}33$^\circ$ \\
600 & \phantom{$^\circ$}175$^\circ$ & \phantom{$^\circ$}490$^\circ$ & \phantom{$^\circ$}1170$^\circ$ & \phantom{$^\circ$}9$^\circ$ & \phantom{$^\circ$}30$^\circ$ \\
650 & \phantom{$^\circ$}175$^\circ$ & \phantom{$^\circ$}510$^\circ$ & \phantom{$^\circ$}1450$^\circ$ & \phantom{$^\circ$}9$^\circ$ & \phantom{$^\circ$}28$^\circ$ \\
700 & \phantom{$^\circ$}380$^\circ$ & \phantom{$^\circ$}2340$^\circ$ & - & \phantom{$^\circ$}3$^\circ$ & \phantom{$^\circ$}24$^\circ$ \\
750 & \phantom{$^\circ$}410$^\circ$ & \phantom{$^\circ$}2420$^\circ$ & - & \phantom{$^\circ$}2$^\circ$ & \phantom{$^\circ$}22$^\circ$ \\
\hline\hline
\multicolumn{6}{p{220pt}}{Ti has a FWHM of 3$^\circ$ after the 550 $^\circ$C anneal round, and disappears afterwards}
\end{tabular}
\end{table}

Besides cracking, holes also form in the Pt (Fig. 11), resulting in a loss of electrical conductivity. There is also anchor damage, which compromises the resonator's structural stability. After 600 $^\circ$C, the bottom right anchor of the example resonator with two anchors breaks, and after 650 $^\circ$C, the top right anchor breaks. The resonator is bending down into the side where the anchors broke, but the remaining anchors are able to provide enough support to prevent the device from collapsing.

Structural characterization was performed starting after the 550 $^\circ$C annealing step using a high-resolution Bruker-JV D1 X-ray diffractometer. Double axis diffraction (DAD) mode was used to obtain 2\straighttheta:\textomega\ scans to verify the layers of the stack and to study phases present after each annealing cycle (Fig 13\deleted{.}). DAD rocking curves from the Pt and Ti layers were measured to study the evolution of the films throughout the annealing cycles. Triple-axis diffraction (TAD) mode was employed to measure rocking curves of the symmetric (2\underline{11}0) LN and quantify lattice tilt and mosaicity of the film. The incident x-ray beam is conditioned by a Göbel mirror and a (220) channel-cut silicon crystal, which produces a highly collimated monochromatic beam of Cu K\textsubscript{\textalpha 1} radiation. In TAD, the scattered beam optics is a 4-bounce (220) channel-cut silicon analyzer crystal which provides with ~10” acceptance angle, while DAD uses mechanical slits in front of the detector instead, providing with ~1400” acceptance angle. Extracted FWHM, FW(0.01)M, and FW(0.001)M values are reported in Table IX. \added{The FW(0.01)M and FW(0.001)M parameters capture the shape of the diffraction peak tails, revealing material quality changes that the FWHM cannot capture.}

\begin{table}[!t]
\centering
\caption{Metal Resistivity}
\label{tab:resistivity750}
\setlength{\tabcolsep}{3pt}
\begin{tabular}{
  >{\centering\arraybackslash}m{60pt}
  >{\centering\arraybackslash}m{50pt}
  >{\centering\arraybackslash}m{50pt}
  >{\centering\arraybackslash}m{60pt}
}
\hline\hline
Annealing temperature ($^\circ$C) & $\rho$ ($\mu \Omega \cdot cm$) & Fractional $\Delta \rho$ (\%) & Normalized to bulk Pt $\rho$ \\
\hline
Initial\phantom{*}  & 35.39 & 0 & 3.37 \\
550\phantom{*}  & 48.20 & 36.20 & 4.59 \\
600\phantom{*}  & 43.14 & 21.90 & 4.11 \\
650* & 34.36 & -2.91 & 3.27 \\
700* & 33.34 & -5.79 & 3.18 \\
750\phantom{*}  & \multicolumn{3}{>{\centering\arraybackslash}m{165pt}}{All devices fail due to conducitivty loss} \\
\hline\hline
\multicolumn{4}{p{220pt}}{*Rounds where some resistivity devices fail due to conductivity loss}
\end{tabular}
\end{table}

X-Ray diffraction (XRD) peaks corresponding to single crystal (111) oriented Si, single crystal X-cut LN, Pt, and Ti are observed in the 2\straighttheta:\textomega\ scan after the 550 $^\circ$C annealing step, which are also the expected peaks prior to annealing. After annealing at 600 $^\circ$C additional peaks that correspond to LiNb\textsubscript{3}O\textsubscript{8} phase transformation are observed in the (010) orientation. The LiNb\textsubscript{3}O\textsubscript{8} peaks increase in intensity with each annealing cycle. Phase composition analysis shows $<$ 1\% of LiNb\textsubscript{3}O\textsubscript{8} second phase is present for temperatures up to 650 $^\circ$C for 10 hours. \added{While the Curie temperature for bulk LN is reported as 1200 °C, thin-film LN bonded onto a silicon substrate begins a phase transition at an earlier temperature. More study on the material side will be required to decipher the experimental observation.} After 700 $^\circ$C for 10 hours. an amount of ~1.5\% was present and after 750 $^\circ$C for 1 hour ~1.8\% was present.

DAD rocking of the metals shows no significant changes for the Pt rocking curve up to 650 $^\circ$C with a FWHM of 9$^\circ$. Subsequent annealing steps improve the Pt crystallinity to 3$^\circ$ FWHM after 700 $^\circ$C for 10 hours, and 2$^\circ$ FWHM after 750 $^\circ$C for 1 hour. The Ti rocking curve shows a FWHM of 3$^\circ$ after the 550 $^\circ$C for 10 hrs. annealing step. This peak is no longer observed for the subsequent annealing steps, likely due to the formation of Ti-Pt intermetallics \cite{ucla4}. The resistivity of the electrode metal decreases after anneals beyond 550 $^\circ$C, and even drops below the initial resistivity after the 650 $^\circ$C and 700 $^\circ$C anneal (Table X). After the 650 $^\circ$C anneal there is a significant -20\% drop in resistivity. However, the resistivity structures also start becoming open-circuited after the 650 $^\circ$C anneal, which results in less data for the averaged resistivity calculation. After the 750 $^\circ$C anneal, all the resistivity devices become open-circuited. \added{Overall, the initial resistivity decrease from annealing is due to defect redistribution and internal stress relaxation [49]. In subsequent annealing rounds, the resistivity increases due to recrystallization and titanium diffusion [49]. The resistivity decrease after the 600 °C anneal round correlates with the titanium disappearing from the XRD measurement, but the physical cause requires further investigation.} In this high-temperature range, the quality of the metal can improve, but it is prone to losing electrical conductivity, which can render devices inoperable.

TAD rocking curve measurements of the symmetric (2\underline{11}0) LN peak show a FWHM of 172”, FW(0.01)M of 460” and a FW(0.001)\added{M} of 570” after the 550 $^\circ$C annealing rounds. No significant changes are observed in the FWHM \added{value} after the 600 $^\circ$C and 650 $^\circ$C annealing rounds. However, increasing tilt and optimizing annealing parameters, such as implementing slower cooling rates, could help avoid film cracking and preserve device integrity for 10 hours. This increase in tilt misorientation is likely caused by the presence of the second phase, which \replaced{is first}{starts to be} observed after the 600 $^\circ$C annealing round. After 700 $^\circ$C for 10 hrs, a large increase in tilt misorientation, likely caused by cracking of the film, is observed, with a FWHM of 380” and a FW(0.01)M of 2340”. The rocking curve widths remain essentially the same after 750 $^\circ$C for 1 h.

Finally, the small amount (1.8\%) of the second phase formed on this sample does not seem to have a negative impact on device performance, nor does the increase in tilt misorientation observed on the LN film after cracking. Furthermore, optimization of annealing parameters, such as implementing slower cooling rates, could avoid cracking of the film and preserve the integrity of devices.

\subsection{Device Performance}

\begin{figure}[!t]
\centerline{\includesvg[width=\columnwidth]{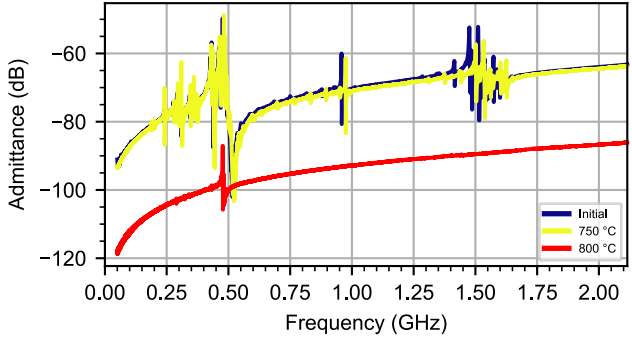}}
\caption{Measured wideband admittance of the example resonator before the initial anneal and after the 750 $^\circ$C anneal.}
\label{fig14}
\end{figure}

\begin{figure}[!t]
\centerline{\includesvg[width=\columnwidth]{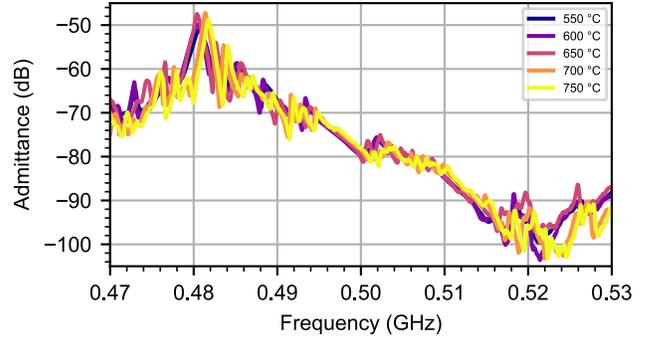}}
\caption{Measured admittance of the example resonator at mode 1 after annealing rounds ranging from 550 $^\circ$C to 750 $^\circ$C.}
\label{fig15}
\end{figure}

\begin{table}[!t]
\centering
\caption{Mode 1 High-Temperature Resonance Parameters}
\label{tab:mode1at750}
\setlength{\tabcolsep}{3pt}
\begin{tabular}{>{\centering\arraybackslash}m{65pt}>{\centering\arraybackslash}m{50pt}>{\centering\arraybackslash}m{30pt}>{\centering\arraybackslash}m{30pt}>{\centering\arraybackslash}m{40pt}}
\hline\hline
Annealing Temperature ($^\circ$C) & Resonant $f$ (MHz) & Fractional $\Delta f$ (\%) & $Q$ & Fractional $\Delta Q$ (\%) \\
\hline
Initial & 467.8 & 0 & 458 & 0 \\
550 & 480.7 & 2.76 & 330 & -28.03 \\
600 & 480.5 & 2.71 & 342 & -25.35 \\
650 & 480.4 & 2.69 & 452 & -1.40 \\
700 & 481.4 & 2.91 & 644 & 40.70 \\
750 & 481.8 & 2.99 & 587 & 28.19 \\
\hline\hline
\end{tabular}
\end{table}

\begin{figure}[!t]
\centerline{\includesvg[width=\columnwidth]{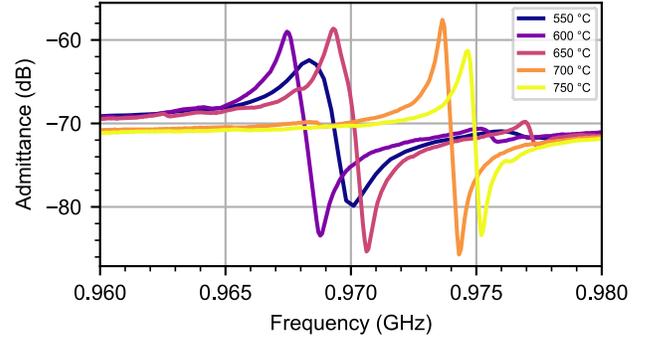}}
\caption{Measured admittance of the example resonator at mode 2 after annealing rounds ranging from 550 $^\circ$C to 750 $^\circ$C.}
\label{fig16}
\end{figure}

\begin{table}[!t]
\centering
\caption{Mode 2 High-Temperature Resonance Parameters}
\label{tab:mode2at750}
\setlength{\tabcolsep}{3pt}
\begin{tabular}{>{\centering\arraybackslash}m{65pt}>{\centering\arraybackslash}m{50pt}>{\centering\arraybackslash}m{30pt}>{\centering\arraybackslash}m{30pt}>{\centering\arraybackslash}m{40pt}}
\hline\hline
Annealing Temperature ($^\circ$C) & Resonant $f$ (MHz) & Fractional $\Delta f$ (\%) & $Q$ & Fractional $\Delta Q$ (\%) \\
\hline
Initial & 958.1 & 0 & 1398 & 0 \\
550 & 968.7 & 1.11 & 851 & -39.14 \\
600 & 967.6 & 0.99 & 1542 & 10.27 \\
650 & 969.5 & 1.19 & 1415 & 1.20 \\
700 & 973.7 & 1.63 & 3285 & 134.93 \\
750 & 974.7 & 1.73 & 2717 & 94.26 \\
\hline\hline
\end{tabular}
\end{table}

\begin{figure}[!t]
\centerline{\includesvg[width=\columnwidth]{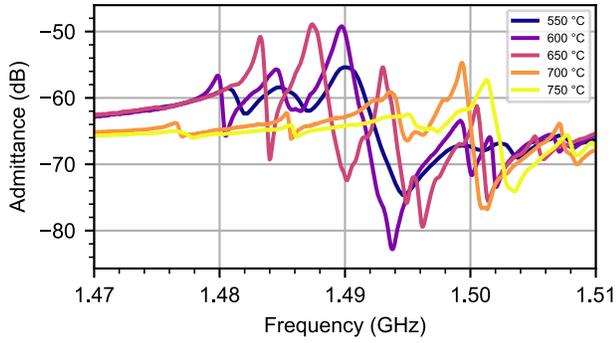}}
\caption{Measured admittance of the example resonator at mode 3 after annealing rounds ranging from 550 $^\circ$C to 750 $^\circ$C.}
\label{fig17}
\end{figure}

\begin{table}[!t]
\centering
\caption{Mode 3 High-Temperature Resonance Parameters}
\label{tab:mode3at750}
\setlength{\tabcolsep}{3pt}
\begin{tabular}{>{\centering\arraybackslash}m{65pt}>{\centering\arraybackslash}m{50pt}>{\centering\arraybackslash}m{30pt}>{\centering\arraybackslash}m{30pt}>{\centering\arraybackslash}m{40pt}}
\hline\hline
Annealing Temperature ($^\circ$C) & Resonant $f$ (GHz) & Fractional $\Delta f$ (\%) & $Q$ & Fractional $\Delta Q$ (\%) \\
\hline
Initial & 1.475 & 0 & 717 & 0 \\
550 & 1.490 & 1.02 & 631 & -11.98 \\
600 & 1.490 & 1.02 & 1371 & 91.20 \\
650 & 1.493 & 1.22 & 1936 & 169.94 \\
700 & 1.499 & 1.63 & 1862 & 159.69 \\
750 & 1.501 & 1.76 & 1243 & 73.32 \\
\hline\hline
\end{tabular}
\end{table}

Overall, the resonant characteristics follow the same trend as earlier annealing rounds, with the exception of a resonant frequency downshift, and the \emph{Q} experiences significant improvements in the late annealing rounds. The wideband admittance of the example resonator is shown in Fig. 14. The admittances of the individual modes are shown in Figs. 15-17, and the extracted resonance parameters are given in Tables XI-XIII. After the 600 $^\circ$C anneal, the resonant frequency is lower compared to the post-550 $^\circ$C anneal round resonant frequency. This is the same annealing round where the Ti disappears from the DAD rocking cruves. In modes 2 and 3, the resonant frequency increases after a 650 $^\circ$C anneal and subsequent anneal rounds, but mode 1's resonant frequency only starts increasing after the 700 $^\circ$C anneal. The overall increase in resonant frequency across the modes can be attributed to increased stress and strain after annealing\added{, which increases the speed of sound in the LN. Since the acoustic wavelength remains fixed due to the interdigitated electrode dimensions, the resonant frequency increases.} After the 700 $^\circ$C anneal, all modes exhibit \emph{Q}s larger than their initial value. Modes 2 and 3 have significant \emph{Q} improvements of over 100\% compared to their original value after the 700 $^\circ$C anneal, as well as after the 650 $^\circ$C anneal for mode 2. In addition to the metal resistivity decreasing for functional devices, some of the resonator's anchors break (Fig. 10). Less electrical energy is lost because of lower resistivity, and less mechanical energy is lost because the resonator can vibrate more freely compared to when all the anchors are intact. \added{Overall, the changes in \emph{Q} do not directly correlate with the change in metal resistivity, suggesting a significant \emph{Q} contribution from the LN’s elasticity as it changes after annealing rounds.} The origins of the frequency drifting and \emph{Q} changes are quite intriguing for future work, but are out of the scope for this work on thermal endurance. \added{Future work on quantifying high-temperature damage through multi-physics coupling models could be of interest.}

After the 800 $^\circ$C anneal, the example resonator no longer functions. Although in Fig. 14 a small resonance can be seen around mode 1's frequency, the much lower capacitance indicates broken anchors because only part of the signal gets capacitively fed into the resonator via the cracked point; furthermore, modes 2 and 3 are no longer visible, and the admittance drops significantly compared to previous annealing rounds. Out of 60 measured resonators, 33 (55\%) became inoperable after the 700 $^\circ$C anneal (Table XIV). This suggests the Si cracking, which also occurs after the 700 $^\circ$C anneal, is a major failure point of the devices. Prior to the 700 $^\circ$C anneal, the only devices that fail are anchored devices. Additionally, 20 resonators had anchors, while 40 resonators had no anchors. Although some resonators are able to operate after a 750 $^\circ$C anneal, it is not near LN's Curie temperature of 1200 $^\circ$C. \added{The device performance gradually degrades at higher temperatures due to a combination of electrode and piezoelectric layer degradation, and also structural damage due to thermal mismatch.} Improvements can be made to the anchor design for enhanced stability, the metal design to delay metal failure, and different materials can be considered to better match the CTE of LN. \added{Future studies can focus on identifying stack and device configurations for better thermal endurance.} \added{Future broader works on higher temperature systems, e.g., via developing higher temperature piezoelectric and electrode materials, would be of research interest, as well as advanced characterization and modeling methods.} Nevertheless, functionality after a 750 $^\circ$C anneal demonstrates the suspended thin-film LN platform's viability for high-temperature applications.

\begin{table}[!t]
\centering
\caption{Resonator Failure Points}
\label{tab:resonatorfailure}
\setlength{\tabcolsep}{3pt}
\begin{tabular}{
  >{\centering\arraybackslash}m{60pt}
  >{\centering\arraybackslash}m{50pt}
  >{\centering\arraybackslash}m{50pt}
}
\hline\hline
Annealing Temperature ($^\circ$C) & Failed devices after anneal & Total failed devices \\
\hline
550 & 1 & 1 \\
600 & 1 & 2 \\
650 & 2 & 4 \\
700 & 33 & 37 \\
750 & 2 & 39 \\
\hline\hline
\end{tabular}
\end{table}

\begin{figure}[!t]
\centerline{\includesvg[width=\columnwidth]{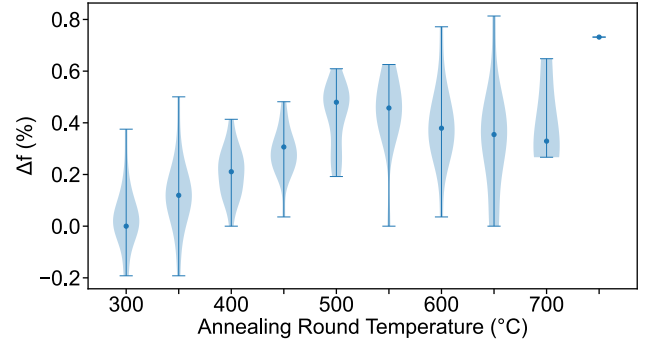}}
\caption{Distributions of percent change in resonant frequency shift relative to the 250 °C anneal round for S0 modes near 500 MHz. The median of each distribution is depicted with a circle. The frequency shifts are normalized to the 250 °C anneal round due to the large variance of frequency shifts after the first anneal. The frequencies of subsequent anneals stay within one percent of the resonant frequency after the initial anneal. After the 550 °C and following anneal rounds, there are fewer points in the distributions due to device failure.}
\label{fig18}
\end{figure}

\section{Conclusion}
\replaced{This paper quantified the thermal endurance of suspended thin-film X-cut stoichiometric LN by characterizing Lamb wave S0 mode acoustic resonators and meandering line resistivity structures with a Pt/Ti metal stack all on the same substrate. The devices were subjected to incremental annealing rounds from 250 $^\circ$C to 800 $^\circ$C. The resonators experience an overall increase in resonant frequency, maintaining a frequency within 1\% of their initial anneal value. After the 600 $^\circ$C anneal, the Ti disappears from XRD measurements and a LiNb\textsubscript{3}O\textsubscript{8} phase forms. After the 650 $^\circ$C anneal, metal discontinuity from hole formation was observed. After the 700 $^\circ$C anneal, many devices failed due to LN cracking and delamination. Despite these damages, the survival of some resonators after a 750 $^\circ$C anneal demonstrates the platform's strong thermal endurance. These results provide baseline data to guide robust device and stack choices for MEMS and related LN systems operating in harsh thermal environments.}{This paper reports a suspended thin-film LN acoustic resonator among 60 in a high-temperature study surviving annealing temperatures of up to 750 $^\circ$C, marking the highest temperature point the platform has been demonstrated in. Despite thermal damage, including cracking, loss of conductivity, and lower LN quality, the survival of the resonator demonstrates the strong thermal endurance of the platform. The platform can be used for MEMS in harsh thermal environments and can push other suspended LN systems, such as optical systems, into higher temperature applications.}

\section*{Acknowledgment}
The authors would like to thank Dr. Todd Bauer for the helpful discussion.


\begin{thebibliography}{00}

\bibitem{transducers} M. Chaudhari, N. Ahmed, V. Tallavajhula, J. Campbell, Y. Wang, Z. Du, and R. Lu, “Thermal Resilience of Suspended Thin-Film Lithium Niobate Acoustic Resonators up to 550~\textdegree C,” \textit{Transducers}, Orlando, FL, USA, Jun.-Jul. 2025, pp. 1859-1862, doi: \href{https://doi.org/10.1109/Transducers61432.2025.11109295}{10.1109/Transducers61432.2025.11109295}.

\bibitem{background1} H.-D. Ngo \emph{et al.}, ``The roadmap for development of piezoresistive micro mechanical sensors for harsh environment applications,'' in \emph{ICST}, Sydney, NSW, Australia, Dec. 2017, pp. 1-6, doi: \href{https://doi.org/10.1109/icsenst.2017.8304457}{10.1109/icsenst.2017.8304457}.

\bibitem{background2} Y. Kim and H. Y. Choi, ``Development of High-Temperature Position Sensors for Control of Actuators in Aerospace Systems,'' in \emph{ICMAE}, Budapest, Jul. 2018. pp. 6-10, doi: \href{https://doi.org/10.1109/icmae.2018.8467645}{10.1109/icmae.2018.8467645}.

\bibitem{background3} Y. Seo, D. Kim, and N. A. Hall, ``High-Temperature Pieozoelectric Pressure Sensors for Hypersonic Flow Measurements,'' in \emph{Transducers \& Eurosensors XXXIII}, Berlin, Germany, Jun. 2019, pp. 2110-2113, doi: \href{https://doi.org/10.1109/transducers.2019.8808755}{10.1109/transducers.2019.8808755}.

\bibitem{sisensor1} J. Ren, M. Ward, P. Kinnell, R. Craddock, and X. Wei, ``Plastic Deformation of Micromachined Silicon Diaphragms with a Sealed Cavity at High Temperatures,'' \emph{Sensors}, vol. 16, no. 2, p. 204, Feb. 2016, doi: \href{https://doi.org/10.3390/s16020204}{10.3390/s16020204}.

\bibitem{sisensor2} Z. Yao \emph{et al.}, ``A High-Temperature Piezoresistive Pressure Sensor with an Integrated Signal-Conditioning Circuit,'' \emph{Sensors}, vol. 16, no. 6, p. 913, Jun. 2016, doi: \href{https://doi.org/10.3390/s16060913}{10.3390/s16060913}.

\bibitem{opticsensor1} Y. Chen \emph{et al.}, ``Recent Progress in MEMS Fiber-Optic Fabry-Perot Pressure Sensors,'' \emph{Sensors}, vol. 24, no. 4, p. 1079, Feb. 2024, doi: \href{https://doi.org/10.3390/s24041079}{10.3390/s24041079}.

\bibitem{materialln1} A. Widenfelder, J. Shi, P. Fielitz, G. Borchardt, K. D. Becker, and H. Fritze, ``Electrical and electromechanical properties of stoichiometric lithium niobate at high-temperatures,'' \emph{Solid State Ionics}, vol. 225, pp. 26-29, Oct. 2012, doi: \href{https://doi.org/10.1016/j.ssi.2012.02.026}{10.1016/j.ssi.2012.02.026}.

\bibitem{materialln2} L. C. Sauze \emph{et al.}, ``Effect of the annealing treatment on the physical and structural properties of LiNbO\textsubscript{3} thin films deposited by radio-frequency sputtering at room temperature,'' \emph{Thin Solid Films}, vol. 726, p. 138660, May 2021, doi: \href{https://doi.org/10.1016/j.tsf.2021.138660}{10.1016/j.tsf.2021.138660}.

\bibitem{materialln3} S. Hurskyy \emph{et al.}, ``Electrical properties and temperature stability of Li-deficient and near stoichiometric Li(Nb,Ta)O\textsubscript{3} solid solutions up to 900 $^\circ$C,'' \emph{Solid State Ionics}, vol. 399, p. 116285, Oct. 2023, doi: \href{https://doi.org/10.1016/j.ssi.2023.116285}{10.1016/j.ssi.2023.116285}.

\bibitem{pzt1} T. Yanagitani, K. Katada, M. Suzuki, and K. Wasa, ``High electromechanical coupling in PZT expitaxial thick film resonators at 550 $^\circ$C,'' in \emph{IFCS}, Taipei, Taiwan, May 2014, pp. 1-3, doi: \href{https://doi.org/10.1109/fcs.2014.6859872}{10.1109/fcs.2014.6859872}.

\bibitem{pzt2} T. Kibe, T. Kaneko, and M. Kobayashi, ``High Temperature Performance of PbTiO\textsubscript{3}/PZT Ultrasonic Transducer above 400 $^\circ$C'', in \emph{IUS}, Taipei, Taiwan, Oct. 2015, pp. 1-4, doi: \href{https://doi.org/10.1109/ultsym.2015.0164}{10.1109/ultsym.2015.0164}.

\bibitem{pzt3} M. Asadnia, A. G. P. Kottapalli, J. M. Miao, A. B. Randles, and J. M. Tsai, ``Performance analysis of PZT (0.52/0.48) for high temperature and pressure sensing applications,'' in \emph{Transducers \& Eurosensors XXVII)}, Barcelona, Spain, Jun. 2013, pp. 984-987, doi: \href{https://doi.org/10.1109/transducers.2013.6626934}{10.1109/transducers.2013.6626934}.

\bibitem{aln1} C.-M. Lin, T.-T. Yen, V. V. Felmetsger, M. A. Hopcroft, J. H. Kuypers, and A. P. Pisano, ``Thermal compensation for aluminum nitride Lamb wave resonators operating at high temperature,'' in \emph{IFCS}, Newport Beach, CA, USA: IEEE, Jun. 2010, pp. 14-18, doi: \href{https://doi.org/10.1109/freq.2010.5556381}{10.1109/freq.2010.5556381}.

\bibitem{aln2} Y.-J. Lai, W.-C. Li, C.-M. Lin, V. V. Felmetsger, and A. P. Pisano, ``High-temperature stable piezoelectric aluminum nitride energy harvesters utilizing elastically supported diaphragms,'' in \emph{Transducers \& Eurosensors XXVII}, Barcelona, Spain, Jun. 2013, pp. 2268-2271, doi: \href{https://doi.org/10.1109/transducers.2013.6627257}{10.1109/transducers.2013.6627257}.

\bibitem{aln3} F. T. Goericke, M. W. Chan, G. Vigevani, I. Izyumin, B. E. Boser, and A. P. Pisano, ``High temperature compatible aluminum nitride resonating strain sensor,'' in \emph{Transducers}, Beijing, China, June 2011, pp. 1994-1997, doi: \href{https://doi.org/10.1109/transducers.2011.5969198}{10.1109/transducers.2011.5969198}.

\bibitem{alscn1} J. Wang, M. Park, and A. Ansari, ``Thermal Characterization of Ferroelectric Aluminum Scandium Nitride Acoustic Resonators,'' in \emph{MEMS}, Gainesville, FL, USA, Jan. 2021, pp. 214-217, doi: \href{https://doi.org/10.1109/mems51782.2021.9375203}{10.1109/mems51782.2021.9375203}.

\bibitem{alscn2} M. Li, B. Chen, J. Xie, W. Song, and Y. Zhu, ``Effects of post-annealing on texture evolution of sputtered ScAlN films,'' in \emph{IUS}, Las Vegas, NV, USA, Sep. 2020, pp. 1-3, doi: \href{https://doi.org/10.1109/ius46767.2020.9251741}{10.1109/ius46767.2020.9251741}.

\bibitem{alscn3} V. Gaddam \emph{et al.}, ``Aluminum Scandium Nitride as a Functional Material at 1000 $^\circ$C,'' \emph{Advanced Electronic Materials}, vol. 11, no. 6, Mar. 2025, doi: \href{https://doi.org/10.1002/aelm.202400849}{10.1002/aelm.202400849}.

\bibitem{ln1} J. Streque \emph{et al.}, ``Design and Characterization of High-Q SAW Resonators Based on the AlN/Sapphire Structure Intended for High-Temperature Wireless Sensor Applications,'' \emph{IEEE Sensors J.}, vol. 20, no. 13, pp. 6985-6991, Jul. 2020, doi: \href{https://doi.org/10.1109/jsen.2020.2978179}{10.1109/jsen.2020.2978179}.

\bibitem{ln2} T. Aubert \emph{et al.}, ``First investigations on stoichiometric lithium niobate as piezoelectric substrate for high-temperature surface acoustic waves applications,'' in \emph{IEEE SENSORS}, Glasgow, Oct. 2017, pp. 1-3, doi: \href{https://doi.org/10.1109/icsens.2017.8234068}{10.1109/icsens.2017.8234068}.

\bibitem{ln3} J. H. Y. Siu, L. Hoff, and M. Frijlink, ``High Temperature Performance of 3 MHz 36$^\circ$ Y-Cut Lithium Niobate Ultrasonic Transducer for Non-Destructive Testing at 550 $^\circ$C,'' in \emph{IUS}, Montreal, QC, Canada, Sep. 2023, pp. 1-4, doi: \href{https://doi.org/10.1109/ius51837.2023.10308184}{10.1109/ius51837.2023.10308184}.

\bibitem{platformcompare} P. Muralit, ``Which is the best thin film piezoelectric material?,'' \emph{IUS}, Washington, DC, USA, 2017, pp. 1-3, doi: \href{https://doi.org/10.1109/ULTSYM.2017.8092807}{10.1109/ULTSYM.2017.8092807}.

\bibitem{ln4} H. Weng, F. L. Duan, Y. Zhang, and M. Hu, ``High Temperature SAW Sensors on LiNbO\textsubscript{3} Substrate With SiO\textsubscript{2} Passivation Layer,'' \emph{IEEE Sensors J.}, vol. 19, no. 24, pp. 11814-11818, Dec. 2019, doi: \href{https://doi.org/10.1109/jsen.2019.2937935}{10.1109/jsen.2019.2937935}.

\bibitem{ln5} J. Streque \emph{et al.}, ``Stoichiometric Lithium Niobate Crystals: Towards Identifiable Wireless Surface Acoustic Wave Sensors Operable up to 600 $^\circ$C,'' \emph{IEEE Sens. Lett.}, vol. 3, no. 4, pp. 1-4, Apr. 2019, doi: \href{https://doi.org/10.1109/lsens.2019.2908691}{10.1109/lsens.2019.2908691}.

\bibitem{ln6} A. O. Ghoname, A. E. Hassanien, E. Chow, and S. Gong, ``Harsh Environment SAW Resonators on Thin Film Lithium Niobate Substrate for RF Wireless Sensing Applications,'' in \emph{UFFC-JS}, Taipei, Taiwan, Sep. 2024, pp. 1-4, doi: \href{https://doi.org/10.1109/uffc-js60046.2024.10793803}{10.1109/uffc-js60046.2024.10793803}.

\bibitem{ln7} W. Gubinelli \emph{et al.}, ``Microacoustic Shear Horizontal Lithium Niobate Leaky SAWs for Harsh Environment Sensing,'' in \emph{ IEEE Sensors}, Kobe, Japan: IEEE, Oct. 2024, pp. 1-4, doi: \href{https://doi.org/10.1109/sensors60989.2024.10785043}{10.1109/sensors60989.2024.10785043}.

\bibitem{ln8} J. Maufay \emph{et al.}, ``Pushing the Limits of LiNbO\textsubscript{3}-based High Temperature SAW Sensors,'' in \emph{2021 IEEE Sensors}, Sydney, Australia: IEEE, Oct. 2021, pp. 1-4, doi: \href{https://doi.org/10.1109/sensors47087.2021.9639644}{10.1109/sensors47087.2021.9639644}.

\bibitem{ln9} T.-H. Hsu, K.-J. Tseng, and M.-H. Li, ``Wideband and High Quality Factor Shear Horizontal SAW Resonators with Improved Temperature Stability in LNOI Platform,'' in \emph{IFCS-EFTF}, Gainesville, FL, USA, Jul. 2021, pp. 1-4, doi: \href{https://doi.org/10.1109/eftf/ifcs52194.2021.9604308}{10.1109/eftf/ifcs52194.2021.9604308}.

\bibitem{susln} S. R. Eisner, C. A. Chapin, R. Lu, Y. Yang, S. Gong, and D. G. Senesky, ``A Laterally Vibrating Lithium Niobate MEMS Resonator Array Operating at 500 $^\circ$C in Air,'' \emph{Sensors}, vol. 21, no. 1, p. 149, Dec. 2020, doi: \href{https://doi.org/10.3390/s21010149}{10.3390/s21010149}.

\bibitem{xcutapp1} R. Tetro, L. Colombo, W. Gubinelli, G. Giribaldi, and M. Rinaldi, ``X-Cut Lithium Niobate S0 Mode Resonators for 5G Applications,'' \emph{MEMS}, Austin, TX, USA, 2024, pp. 1102-1105, doi: \href{https://doi.org/10.1109/MEMS58180.2024.10439491}{10.1109/MEMS58180.2024.10439491}.

\bibitem{xcutapp2} F. Hartmann, S. E. Kuçuk Eroglu, E. Navarro-Gessé, C. Collado, J. Mateu and L. G. Villanueva, ``A 5G n77 Filter Using Shear Bulk Mode Rsonator With Crystalline X-cut Lithium Niobate Films,'' \emph{IMFW}, Cocoa Beach, FL, USA, 2024, pp. 78-80, doi: \href{https://doi.org/10.1109/IMFW59690.2024.10477114}{10.1109/IMFW59690.2024.10477114}.

\bibitem{xcutapp3} L. Shao \emph{et al.}, ``Microwave-to-optical conversion using lithium niobate thin-film acoustic resonators,'' \emph{Optica}, vol. 6, no. 12, pp. 1498-1505, Dec. 2019, doi: \href{https://doi.org/10.1364/OPTICA.6.001498}{10.1364/OPTICA.6.001498}.

\bibitem{xcutapp4} L. Shao \emph{et al.}, ``Integrated microwave acousto-optic frequency shifter on thin-film lithium niobate,'' \emph{Optics Express}, vol. 28, no. 16, pp. 23728-23738, Aug. 2020, doi: \href{https://doi.org/10.1364/OE.397138}{10.1364/OE.397138}.

\bibitem{xcutapp5} R. Cheng \emph{et al.}, ``Frequency comb generation via synchronous pumped χ\textsuperscript{(3)} resonator on thin-film lithium niobate,'' \emph{Nautre Communications}, vol. 15, May 2024, doi: \href{https://doi.org/10.1038/s41467-024-48222-3}{10.1038/s41467-024-48222-3}.

\bibitem{metaltable1} ``Metals and Alloys - Melting Temperatures.'' The Engineering ToolBox. Accessed: Jul. 25, 2025. [Online.] Available: \href{https://www.engineeringtoolbox.com/melting-temperature-metals-d_860.html}{https://www.engineeringtoolbox.com/melting-temperature-metals-d\_860.html}.

\bibitem{metaltable2} ``Resistivity and Conductivity - Temperature Coefficients Common Materials.'' The Engineering ToolBox. Accessed: Jul. 25, 2025. [Online.] Available: \href{https://www.engineeringtoolbox.com/resistivity-conductivity-d_418.html}{https://www.engineeringtoolbox.com/resistivity-conductivity-d\_418.html}.

\bibitem{pt1} E. Çiftyürek, K. Sabolsky, and E. M. Sabolsky, ``Platinum thin film electrodes for high-temperature chemical sensor applications,'' \emph{Sensors and Actuators B: Chemical}, vol. 181, pp. 702-714, May 2013, doi: \href{https://doi.org/10.1016/j.snb.2013.02.058}{10.1016/j.snb.2013.02.058}.

\bibitem{pt2} W. Chen, P. Wang, Q. Cui, Z. Qiang, L. Qiao, and Q. Li, ``Effect of titanium adhesion layer on the thermal stability of platinum films during vacuum high temperature treatment,'' \emph{Vacuum}, vol. 226, p. 113295, Aug. 2024, doi: \href{https://doi.org/10.1016/j.vacuum.2024.113295}{10.1016/j.vacuum.2024.113295}.

\bibitem{pt3} F. Yi, W. Osborn, J. Betz, and D. A. LaVan, ``Interactions of Adhesion Materials and Annealing Environment on Resistance and Stability of MEMS Platinum Heaters and Temperature Sensors,'' \emph{J. Microelectromech. Syst.}, vol. 24, no. 4, pp. 1185-1192, Aug. 2015, doi: \href{https://doi.org/10.1109/jmems.2015.2394483}{10.1109/jmems.2015.2394483}.

\bibitem{tibetter} V. Guarnieri, L. Biazi, R. Marchiori, and A. Lago, ``Platinum metallization for MEMS application: Focus on coating adhesion for biomedical applications,'' \emph{Biomatter}, vol. 4, no. 1, Jan. 2014, doi: \href{https://doi.org/10.4161/biom.28822}{10.4161/biom.28822}.

\bibitem{lwr} R. Lu and S. Gong, ``RF acoustic microsystems based on suspended lithium niobate thin films: advances and outlook,'' \emph{J. Micromech. Microeng.}, vol. 31, no. 11, p. 114001, Nov. 2021, doi: \href{https://doi.org/10.1088/1361-6439/ac288f}{10.1088/1361-6439/ac288f}.

\bibitem{thermalshock} N. Miyazaki, A. Hattori, M. Refinery, and H. Uchida, ``Thermal shock cracking of lithium niobate single crystal,'' \emph{J. Mater. Sci.: Mater. Electron.}, vol. 8, pp. 133-138, Jun. 1997, doi: \href{https://doi.org/10.1023/A:1018581710568}{10.1023/A:1018581710568}.

\bibitem{pyroelectric} J. Biendl, F. Dreher, M. Protte, J. P. Höpker, V. B. Verma, and T. J. Bartley, ``Overcoming Pyroelectricity to Improve Integrated Superconducting Detector Fabrication Yield on Lithium Niobate,'' \emph{CLEO}, Charlotte, NC, USA, 2024, doi: \href{https://doi.org/10.1364/CLEO_AT.2024.JW2A.100}{10.1364/CLEO\_AT.2024.JW2A.100}.

\bibitem{ucla1} H. Takagi, R. Maeda, N. Hosoda, and T. Suga, ``Room-temperature bonding of lithium niobate and silicon wafers by argon-beam surface activation,'' \emph{Appl. Phys. Lett.}, vol. 74, no. 16, pp. 2387-2389, Apr. 1999, doi: \href{https://doi.org/10.1063/1.123860}{10.1063/1.123860}.

\bibitem{ucla2} Y. Okada, Y. Tokumaru, ``Precise determination of lattice parameter and thermal expansion coefficient of silicon between 300 and 1500 K,'' \emph{J. Appl. Phys.}, vol 56, no. 2, pp. 314-320, Jul. 1984, doi: \href{https://doi.org/10.1063/1.333965}{10.1063/1.333965}.

\bibitem{ucla3} Y. Hirsh, S. Gorfman, D. Sherman, ``Clevage and surface energies of LiNbO\textsubscript{3},'' \emph{Acta Mater.}, vol. 193, pp. 338-349, Jul. 2020, doi: \href{https://doi.org/10.1016/j.actamat.2020.03.046}{10.1016/j.actamat.2020.03.046}.

\bibitem{ucla4} S. Hsieh, D. Beck, T. Matsumoto, B. E. Koel, ``Thermal stability of ultrathin titanium films on a Pt(111) substrate,'' \emph{Thin Solid Films}, vol. 466, no. 1-2, pp. 123-127, pp. 123-127, Nov. 2004, doi: \href{https://doi.org/10.1016/j.tsf.2004.03.041}{10.1016/j.tsf.2004.03.041}.

\bibitem{49} \added{U. Schmid, H. Seidel, ``Influence of thermal annealing on the resistivity of titanium/platinum thin films,'' \emph{J. Vac. Sci. Technol. A}, vol. 24, no. 6, pp. 2139-2146, Nov. 2006, \href{https://doi.org/10.1116/1.2359739}{doi: 10.1116/1.2359739}.}

\end{thebibliography}
\end{document}